\documentclass[fleqn,10pt]{wlscirep}
\usepackage[utf8]{inputenc}
\usepackage[T1]{fontenc}
\usepackage{subcaption}
\usepackage{booktabs}
\usepackage{longtable}
\usepackage{makecell}
\usepackage{framed,multirow}
\usepackage{amssymb}
\usepackage{latexsym}

\title{MINDSETS: Multi-omics Integration with Neuroimaging for Dementia Subtyping and Effective Temporal Study}

\author[1,*]{Salma Hassan}
\author[1]{Dawlat Akaila}
\author[1]{Maryam Arjemandi}
\author[2]{Vijay Papineni}
\author[1]{Mohammad Yaqub}
\affil[1]{Mohamed bin Zayed University of Artificial Intelligence, Abu Dhabi, United Arab Emirates}
\affil[2]{Sheikh Shakhbout Medical City, Abu Dhabi, United Arab Emirates}
\affil[*]{Corresponding author at: salma.hassan@mbzuai.ac.ae}


\begin{abstract}
In the complex realm of cognitive disorders, Alzheimer's disease (AD) and vascular dementia (VaD) are the two most prevalent dementia types, presenting entangled symptoms yet requiring distinct treatment approaches. The crux of effective treatment in slowing neurodegeneration lies in early, accurate diagnosis, as this significantly assists doctors in determining the appropriate course of action. However, current diagnostic practices often delay VaD diagnosis, impeding timely intervention and adversely affecting patient prognosis. This paper presents an innovative multi-omics approach to accurately differentiate AD from VaD, achieving a diagnostic accuracy of 89.25\%. The proposed method segments the longitudinal MRI scans and extracts advanced radiomics features. Subsequently, it synergistically integrates the radiomics features with an ensemble of clinical, cognitive, and genetic data to provide state-of-the-art diagnostic accuracy, setting a new benchmark in classification accuracy on a large public dataset. The paper's primary contribution is proposing a comprehensive methodology utilizing multi-omics data to provide a nuanced understanding of dementia subtypes. Additionally, the paper introduces an interpretable model to enhance clinical decision-making coupled with a novel model architecture for evaluating treatment efficacy. These advancements lay the groundwork for future work not only aimed at improving differential diagnosis but also mitigating and preventing the progression of dementia.
\end{abstract}
\begin{document}

\flushbottom
\maketitle
%
%
\thispagestyle{empty}

\section*{Introduction}
Dementia is a complex brain disorder, affecting 55 million individuals globally with ten million new cases annually, and represents a critical public health challenge \cite{WHO}. This umbrella term encompasses various cognitive disorders, including Alzheimer's disease (AD) and vascular dementia (VaD), which are the most prevalent forms. AD, responsible for 60-80\% of dementia cases, is the sixth leading cause of death in the USA, with its mortality rates rising in contrast to those of heart disease and cancer \cite{eisenmenger2023vascular, CDC}. This trend underscores the urgent need for enhanced research, healthcare strategies, and societal support to address the escalating burden of AD. VaD, constituting 10-15\% of dementia cases, shares symptoms with AD but requires distinct treatment due to its cerebrovascular pathology \cite{rajkumar2020delirium}. It results from cerebral ischemic or hemorrhagic events, leading to cognitive decline that varies with the affected brain regions. Early detection and management of vascular risk factors are critical for preventing VaD progression and improving patient outcomes.

The differential diagnosis between AD and VaD is vital given their unique etiologies and treatment approaches. AD primarily impairs memory and cognitive functions related to the hippocampus and entorhinal cortex, affecting language, navigation, and recognition skills \cite{ammar2020language}. In contrast, VaD affects executive functions early, resulting in difficulties with planning, organization, complex tasks, and mood disturbances \cite{borelli2020causes, hirsch2021expanded}. VaD patients have a more severe speech impairment and score up to 7.7\% less on phonetic fluency tasks than AD patients, \cite{hirsch2021expanded}. The Hachinski Ischemic Score (HIS) is vital in distinguishing between degenerative and vascular dementia origins \cite{kim2014clinical}.

Both AD and VaD have a profound impact on individuals' lives, extending far beyond the physiological symptoms. They affect patients' social, psychological, and economic well-being, but the nature and progression of cognitive decline differ between the two. Understanding the multifaceted nature of this impact emphasizes the critical importance of early diagnosis. Our research aims to bridge the gap in diagnostic capabilities, offering new insights and methodologies to enhance early detection and treatment strategies for AD and VaD.

\noindent\textbf{\\Contributions.} This paper introduces a method for differentiating AD and VaD using multi-omics data. The main contributions are:
{\begin{itemize}
    \item Propose a unified and robust multi-omics method to distinguish between AD and VaD from MRI radiomics, genetic data, and clinical assessments to reveal their pathological differences while achieving state-of-the-art classification performance on a large public dataset.
    \item Introduce an explainable model to facilitate its adoption in clinical practice and provide a personalized interpretable diagnosis.
    \item Develop a model architecture to monitor longitudinal changes in MCI diagnosis confidence and probability to assess treatment efficacy.   
\end{itemize}}
These contributions underscore our paper's potential to transform the diagnostic landscape for AD and VaD, offering a clinically applicable solution to bridge current gaps in the diagnosis and management of dementia.

\section*{Background}
\subsection*{Diagnostic Challenges in AD and VaD}
\subsubsection*{Symptomatic Overlap and Clinical Complications}
The differential diagnosis of AD and VaD remains a formidable challenge in the neurocognitive domain, primarily due to the significant symptomatic overlap between these disorders. Both AD and VaD share common symptoms, such as memory loss, executive dysfunction, and difficulties with daily activities, which complicate clinical assessments \cite{kalaria2018pathology}. This overlap complicates clinical evaluations and underscores the urgent need for precision in diagnosis to enable effective treatments and improve patient prognoses.

\subsubsection*{Limitations of Traditional Diagnostic Tools}
Traditional diagnostic tools, including clinical interviews, cognitive testing, and basic imaging techniques, often fall short in accurately differentiating between AD and VaD \cite{arevalo2014diagnostic}. Cognitive assessments may not sufficiently capture the subtle differences in disease pathology, while standard imaging techniques like CT and conventional MRI may not provide the resolution needed to distinguish between the vascular and neurodegenerative changes characteristic of these conditions \cite{arevalo2014diagnostic}. This inadequacy in current diagnostic methods highlights the necessity for more advanced, nuanced approaches.

\subsubsection*{Importance of Accurate Differentiation}
Accurate differentiation between AD and VaD is crucial for several reasons. First, it ensures that patients receive the most appropriate treatments, which can vary significantly between the two conditions \cite{ni2024classification}. AD treatments often focus on managing symptoms and slowing disease progression, while VaD management emphasizes preventing further vascular damage through aggressive control of cardiovascular risk factors. Second, precise diagnosis aids in better patient management and prognosis, allowing for tailored therapeutic strategies and more informed clinical decisions \cite{ni2024classification}. Finally, it enhances our understanding of the pathophysiology of these dementias, contributing to the development of more effective interventions.

\subsection*{Differentiating Factors Between AD and VaD}
AD typically has a gradual and progressive onset, characterized by a slow cognitive and functional decline with early loss of awareness\cite{sokolovivc2023neuropsychological}. In contrast, VaD often presents with sudden onset due to cerebrovascular events, followed by a stepwise progression and focal neurological signs. Regarding attention and executive function, AD and VaD are equally impaired in sustained attention. However, selective attention is more impaired in VaD than AD, which can be attributed to cerebrovascular damage affecting brain regions responsible for executive functions \cite{sokolovivc2023neuropsychological}. Regarding language assessment, semantic fluency is more impaired in AD, with patients struggling with naming and language comprehension tasks. In contrast, individuals with VaD perform worse on phonemic fluency measures, reflecting the impact of cerebrovascular damage on language processing areas \cite{arnaoutoglou2017color}. Finally, regarding memory impact, AD shows greater overall memory impairment than VaD, particularly in episodic memory. Participants with VaD often perform better on episodic memory tests, while individuals with AD show significant deficits in recall memory tasks, which are crucial for diagnosis \cite{sokolovivc2023neuropsychological}.

\subsection*{Vascular and Neurodegenerative Changes}
VaD is primarily caused by tissue injury from ischemia, affecting the brain's white matter and subcortical structures. Commonly impacted areas include the basal ganglia, thalamus, and regions responsible for executive function and processing speed \cite{attems2014overlap}. Ischemic lesions, small vessel disease leading to white matter hyperintensities, lacunar infarcts, and microbleeds are also present. On the other hand, AD is characterized by generalized brain atrophy, particularly in the hippocampus and temporal lobes, with early damage often seen in the hippocampus and entorhinal cortex \cite{iadecola2003converging}. As the disease progresses, the cerebral cortex, involved in thought, perception, and language, is affected, leading to ventricular enlargement and cortical thinning \cite{attems2014overlap}. These distinct differences between the two types impact the treatment approaches. While AD treatment focuses on managing symptoms and slowing progression, VaD treatment emphasizes aggressive management of cardiovascular risk factors to prevent further vascular damage. The life expectancy of VaD is shorter, and its diagnosis often comes too late for effective recuperative treatment. 

\subsection*{Advancements in Machine Learning and Neuroimaging}
Research utilizing the Alzheimer's Disease Neuroimaging Initiative (ADNI) dataset has significantly advanced our understanding of AD and VaD. Studies such as \cite{zheng2019machine} have demonstrated the potential of machine learning algorithms combined with longitudinal structural MRI biomarkers to enhance diagnostic precision. These advancements offer valuable tools for clinicians, aiding in the early diagnosis of these distinct dementia types by identifying subtle brain changes over time. MRI provides detailed images of brain structure, allowing for the identification of dementia-specific brain changes. Studies have shown that MRI can detect subtle neurodegenerative markers such as white matter hyperintensities and lacunes, which are critical for distinguishing between AD and VaD in the early stages of disease progression \cite{chouliaras2023use, ghai2020current, bir2021emerging}.

\subsection*{Machine Learning Techniques}
Machine learning techniques, including Support Vector Machines (SVM), K-nearest neighbors (KNN), and Random Forest (RF) classifiers, have improved diagnostic precision by analyzing complex patterns in neuroimaging and biomarker data. These techniques can identify distinctive brain volume differences and other biomarkers, facilitating more accurate early-stage dementia diagnosis \cite{llorens2020cerebrospinal, jorgensen2020age, zheng2019machine, manouvelou2020differential}.

\subsection*{Multi-omics Approach for Improved Diagnosis}
In light of these developments, a multi-omics diagnostic framework is indispensable for addressing the complex interplay of symptoms in AD and VaD and deepening the understanding of their distinct pathophysiological traits. Therefore, we propose a novel approach, MINDSETS (Multi-omics Integration with Neuroimaging for Dementia Subtyping and Effective Temporal Study), to address this challenge. This approach leverages the integration of genetic, proteomic, and neuroimaging data to create a comprehensive diagnostic model that can more accurately distinguish between AD and VaD. By combining these diverse data sources, MINDSETS aims to improve diagnostic accuracy, enable early detection, and tailor treatment strategies to individual patients' needs.

\begin{figure*}[!ht]
\includegraphics[width=\textwidth]{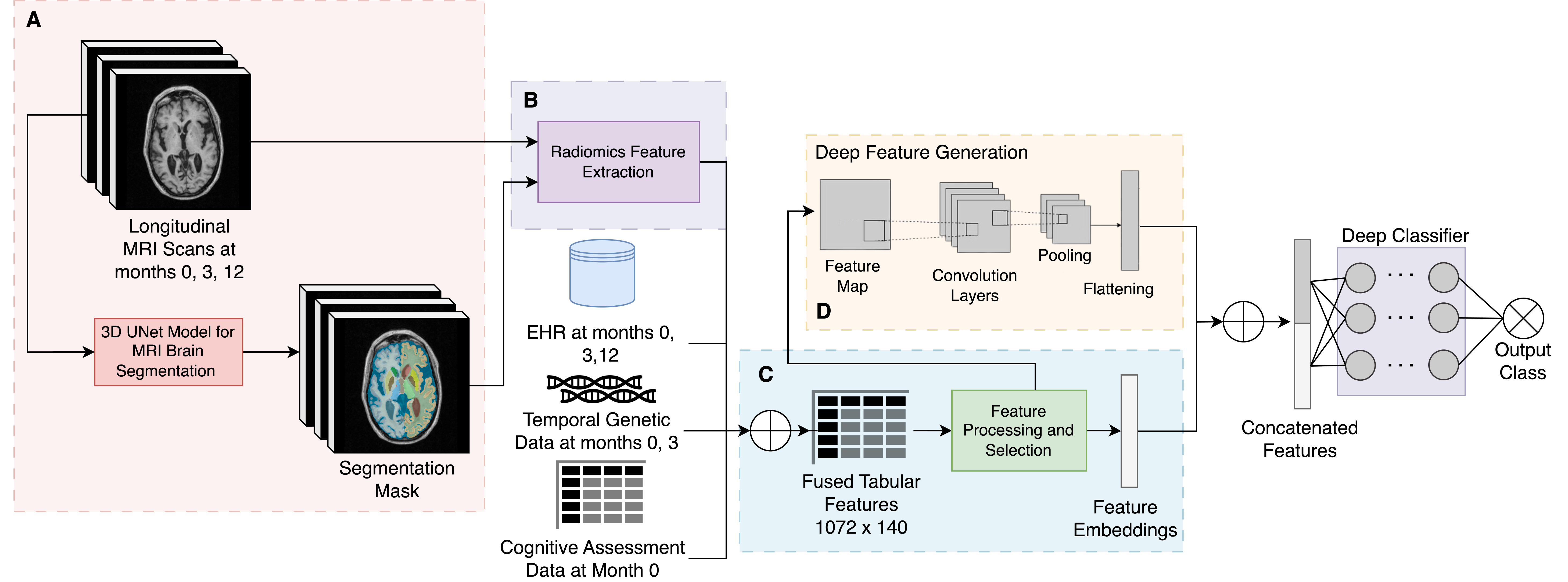}
\caption{The architecture segments the MRI scans to extract radiomics features and fuses them with other multi-omics features. Feature selection is applied, and then features are passed to the Deep Feature Generation module, where discriminative features are generated and concatenated with the raw features.} \label{arch}
\end{figure*}

\subsection*{Proposed MINDSETS Approach}

\begin{figure*}[!ht]
\includegraphics[width=\textwidth]{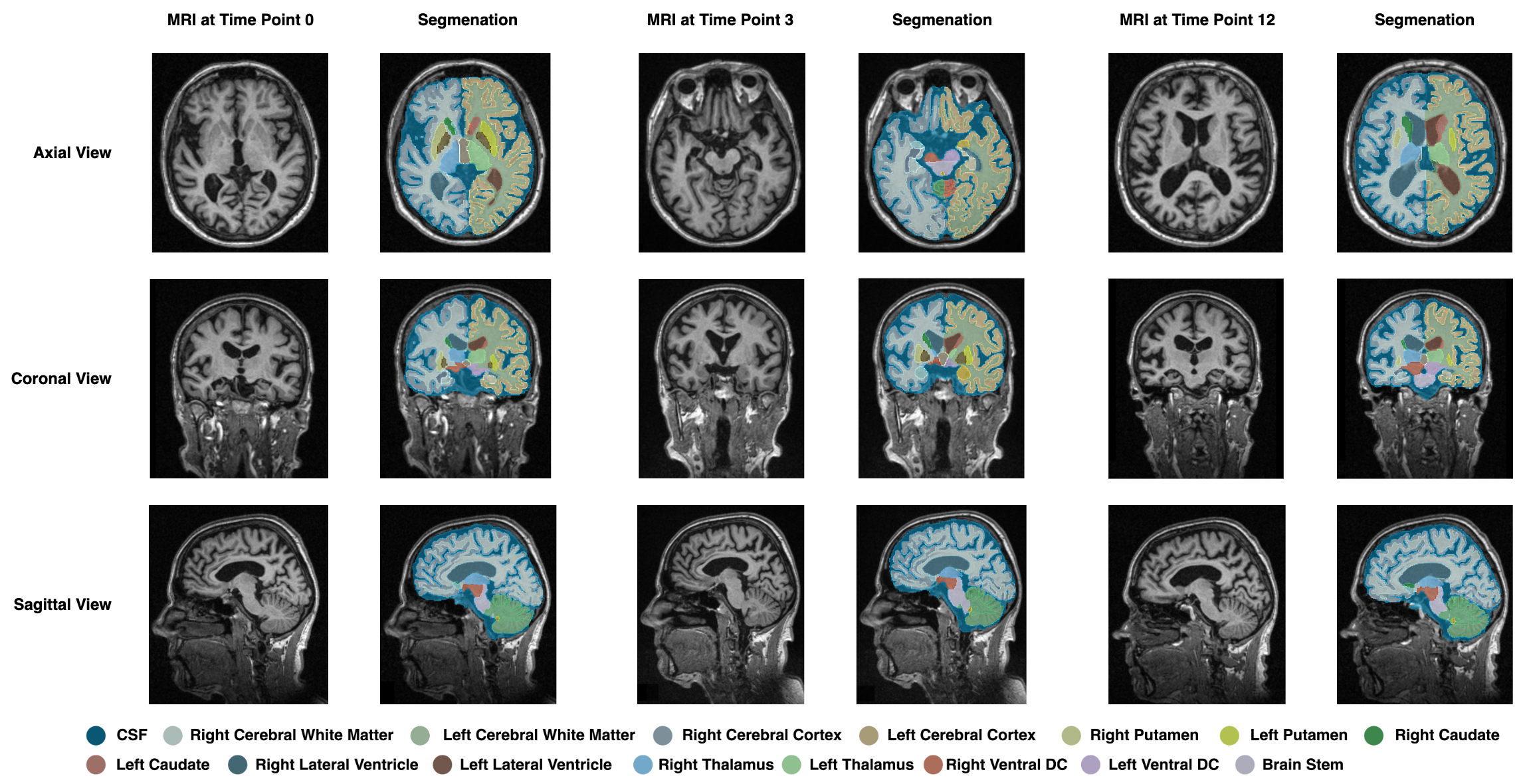}
\caption{Multi-planar MRI Scans with Annotated Segmentation: The image series showcases axial, coronal, and sagittal views of the temporal brain MRIs at time points 0, 3, 12, both in raw form (left column) and with detailed segmentation overlays (right column). The segmentation highlights various brain structures, including the thalamus and cortical regions, using distinct color codes to differentiate between anatomical areas.} \label{mri_temp}
\end{figure*}

\subsubsection*{MRI Scan Segmentation}
The first step is to segment the brain to extract the radiomics features using the mask as illustrated in Figure \ref{arch}A. There are no ground truth segmentation masks available in the dataset. Thus, a pre-trained 3D UNet model was leveraged through SynthSeg, a deep learning tool adept at automatic brain MRI scan segmentation, notable for its robustness to contrast and resolution variations, trained across a diverse set of scans, including those from both elderly diseased and healthy populations \cite{billot2023synthseg}. This model segmented all the temporal MRI scans into 32 different structures as illustrated in Figure \ref{mri_temp}. Three criteria were utilized to validate the accuracy of these segmentation masks. Firstly, SynthSeg's automated quality control mechanism confirmed a high similarity score for all structures when compared to a standard reference. Secondly, comparing the volumes of the segmented structures against known true values revealed minimal discrepancies. Lastly, a manual assessment by an experienced radiologist on a random subset of scans was performed to assess the segmentation performance.

\subsubsection*{Radiomics Features Extraction} The next step was to utilize the segmentation mask for MRI feature extraction, as seen in Figure \ref{arch}B. To facilitate this, PyRadiomics, a Python package for radiomics analysis, was used \cite{van2017computational}. This process involves extracting 137 distinct features from each of the 32 brain structures identified during segmentation, covering features related to shape, first-order statistics, and gray-level attributes. Recent studies underscore the enhanced diagnostic accuracy achieved by using radiomics features, stressing its potential in quantitatively analyzing tissue patterns at the voxel level \cite{yao2023review,szabo2023radiomics}. This approach is particularly pertinent for distinguishing between AD and VaD, as these conditions manifest through specific changes in brain tissues. 

\subsubsection*{Features Fusion and Selection}
Following the extraction of MRI radiomics features, since the output of PyRadiomics is in tabular form, it was concatenated with the tabular clinical, genotype data and the mini-mental state examination (MMSE) cognitive assessment score, as shown in Figure \ref{arch}C. Since AD affects memory more prominently than VaD, the MMSE score was split into memory and processing-related sub-scores. After fusion, various feature selection approaches were compared. These included statistical methods such as chi-squared, principal component analysis, and t-distributed stochastic neighbor embedding. However, FeatureWiz provided the best result by reducing multicollinearity.

\subsubsection*{Deep Feature Generation} 
We propose the Deep Feature Generation (DFG) module to enhance the discriminative power of the architecture and surpass the current diagnostic accuracy, shown in Figure \ref{arch}D. The DFG module utilizes Convolutional Neural Networks (CNNs) to dynamically extract intricate patterns, thus enriching the representation of the data to combine local patterns into new features with deeper insights into the data. The new features are then concatenated with the raw features and passed to a deep classifier to output the class. This process empowers our architecture to capture nuanced relationships and subtle variations within the dataset, consequently improving its ability to discern relevant information for subsequent analysis. Through deep feature generation, we aim to augment the efficacy and robustness of our model, thereby advancing the accuracy and reliability of detention subtype classification.

One of the main aims of this paper is to provide explainability in terms of which features are most significant in deciding for diagnosis. Explainability is especially important in the medical field, where identifying critical features can provide vital insights into the diagnosis, treatment, and prevention of various medical conditions. To address this, several steps were implemented throughout the methodology to steer away from treating the model as a black box. For instance, feature importance scores were calculated during feature selection using techniques such as Random Forest and SHAP (Shapley Additive Explanations). This approach helped identify the most influential features and understand the direction and magnitude of their impact on the diagnostic predictions. Furthermore, visualizations were created to present these feature importance scores, allowing medical professionals to interpret them easily. 

\subsection*{Explainability}
Explainability is especially important in the medical field, where identifying critical features can provide vital insights into the diagnosis, treatment, and prevention of various medical conditions. To address this, several steps were implemented throughout the methodology to avoid treating the model as a black box. For instance, feature importance scores were calculated during feature selection using techniques such as Random Forest and SHAP (Shapley Additive Explanations). This approach helped identify the most influential features and understand the direction and magnitude of their impact on the diagnostic predictions. Furthermore, visualizations were created to present these feature importance scores, allowing medical professionals to interpret them easily.

\section*{Experimental Setup}
\begin{figure}[!ht]
\centering
\includegraphics[width=0.6\textwidth]{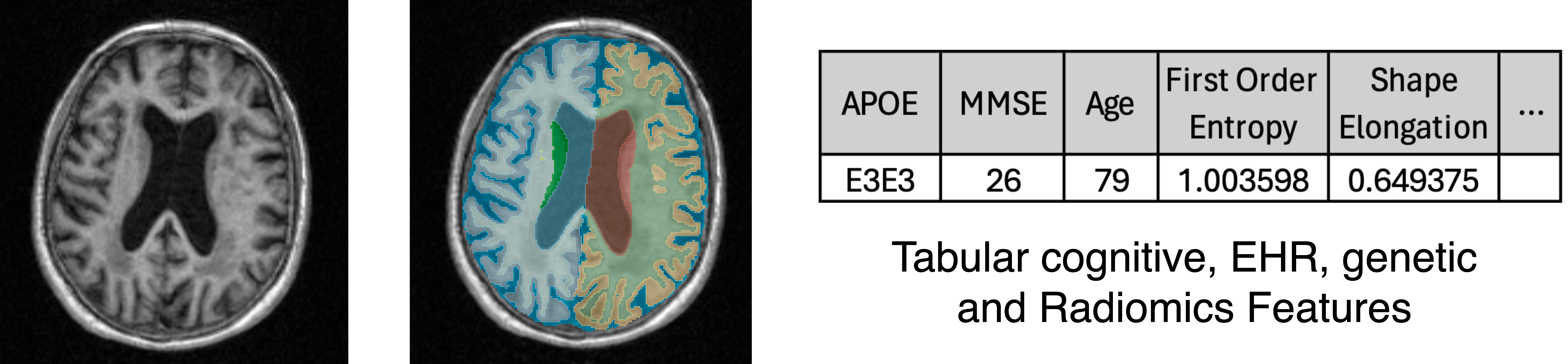}
\caption{An example patient data with MRI, the generated segmentation mask, and a sample set of the tabular data, left to right.} \label{patient_results}
\end{figure}
\subsection*{Dataset} 
Building a large-scale multi-omics dataset targeting dementia subtypes is expensive and time-consuming. There is only one public dataset to our knowledge that offers a rich foundation for dementia research, which is ANMerge \cite{birkenbihl2021anmerge} composed of 1,702 participants. Unlike the ADNI dataset, it encompasses EHR, longitudinal MRI scans, and comprehensive multi-omics data such as genomic profiles; example data is shown in Figure \ref{patient_results}. ANMerge's emphasis on temporal and patient-level data, such as CDR and Camdex, enhances the model's ability to analyze dementia progression and treatment efficacy, which is a unique advantage for personalized analysis compared to ADNI. There are four classes: AD, VaD, Mild Cognitive Impairment (MCI), and Control (CTL). This study leverages the dataset's longitudinal MRI scans to critically discern degenerative versus vascular brain changes. Furthermore, it explores the diagnostic enhancement potential of integrating additional modalities —clinical data, cognitive assessments, and genetic information. Through experimentation, we assess the incremental value of each modality in improving diagnostic accuracy.

\begin{figure*}[!ht]
\centering
\begin{subfigure}{.5\textwidth}
  \centering
  \includegraphics[width=1.0\linewidth]{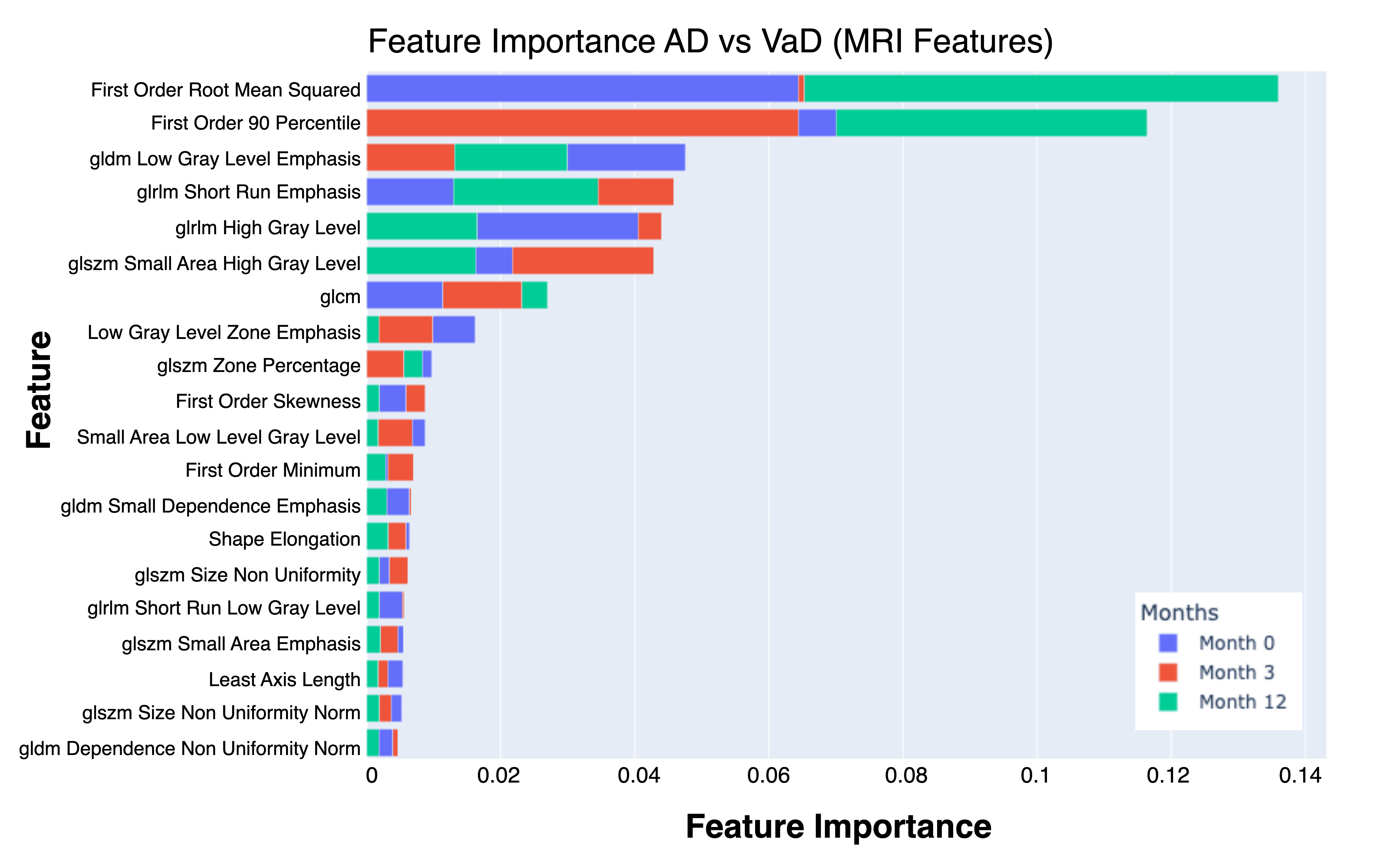}
  \caption{MRI radiomics feature importance}
  \label{fig:notrans}
\end{subfigure}%
\begin{subfigure}{.5\textwidth}
  \centering
  \includegraphics[width=0.97\linewidth]{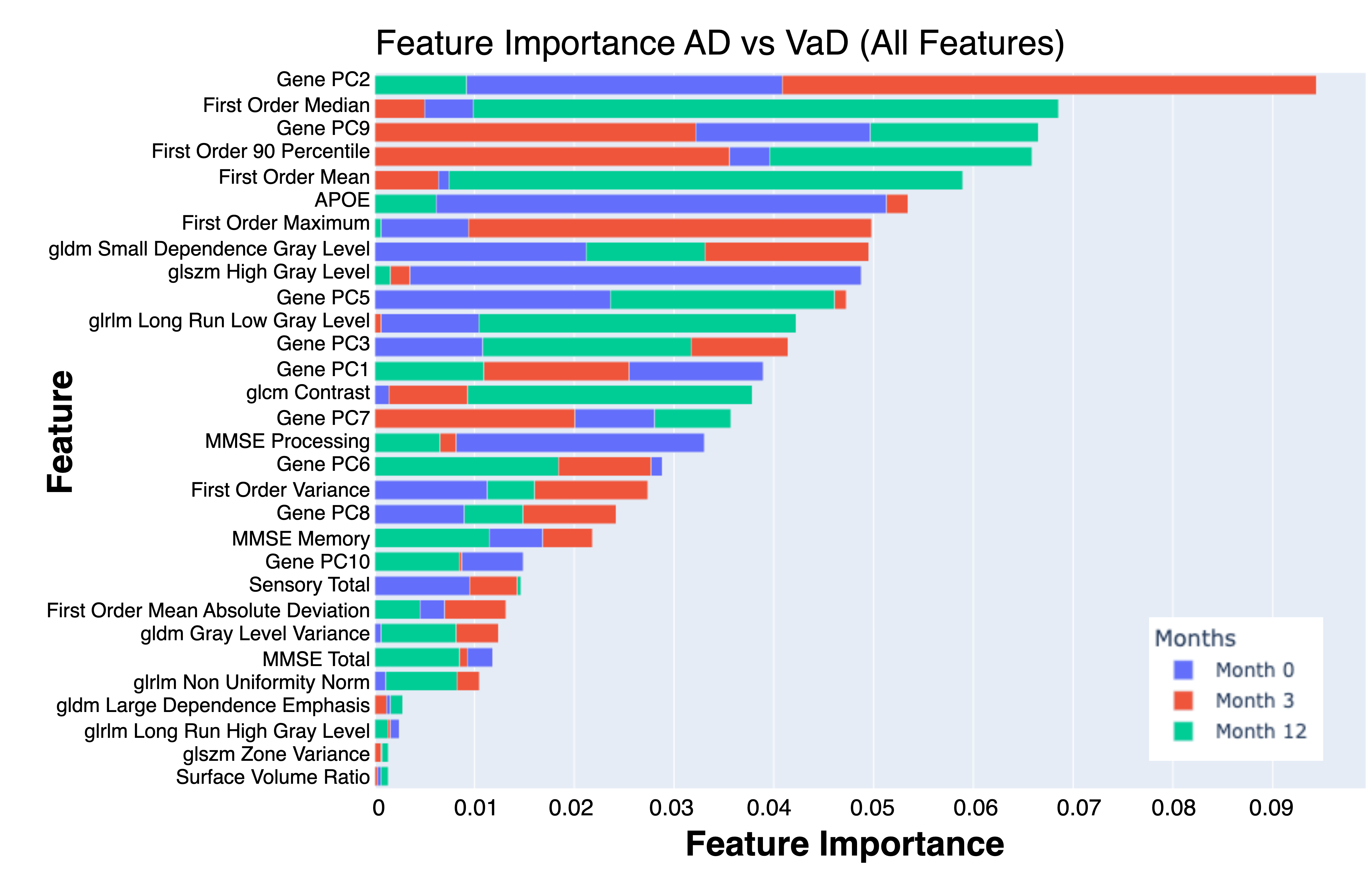}
  \caption{Multi-omics feature importance}
  \label{fig:trans}
\end{subfigure}
\caption{Figure shows the feature importance of MRI radiomics features only versus multi-omics features. Importance is broken down by scan timepoint.}
\label{fig:feature_improtance}
\end{figure*}

\subsection*{Implementation Details} 
For training, a 5 Group K-fold cross-validation strategy was implemented where data was split by patients to avoid leakage. Categorical features were encoded using label encoding, and continuous features were scaled using Standard Scaler. Missing information was imputed with the mean or mode in the case of continuous and categorical, respectively, based on the class group. The hyperparameters used were a maximum of 500 epochs but with early stopping implemented with a patience level of 5. For the DFG module, 14 filters were used with kernel size 7x7. The max pooling layer width was set to 2. The Optuna library was used for hyperparameters tuning \cite{akiba2019optuna}.

\subsection*{Ablation Studies} 
\subsubsection*{3D CNN Model using MRI Only}
The first approach considered for this paper was an alternative to implementing the segmentation and radiomics method. Therefore, a direct image-based 3D CNN model was tested to investigate whether using raw 3D MRI alone could yield results comparable to our proposed methodology. Moreover, several machine learning and deep learning algorithms, such as random forest, support vector machine, multi-layer perception, and deep factorization machines, were implemented to test their performance against the proposed approach. Also, the model's performance was evaluated with and without the DFG module to isolate its contribution to the overall improvement. 

\subsubsection*{Feature Selection}
Various feature selection approaches were implemented to compare. These included statistical methods such as chi-squared, variance threshold, principal component analysis (PCA), t-distributed stochastic neighbor embedding (t-SNE), forward selection, backward elimination, and exhaustive searching. In addition, several libraries, such as mlxtend, were used. However, as mentioned, FeatureWiz provided the best result by working to reduce multicollinearity between the features while maintaining a high correlation with the target. 

\subsubsection*{Model Selection}
Several ML and DL models were investigated to determine which would work best. Lazypredict library was used to quickly assess which ML algorithm would work best before hyperparameter tuning. SVM and random forest were the top two algorithms, respectively, with SVM outperforming RF in all experiments. A simple 1D CNN and tabular ResNet were implemented and tested for the DL approach. ResNet model provided a marginal boost in accuracy in some tests but was inconsistent through all the tests. TabNet library was also used to test its classifier, but it was not the highest performing. Finally, using the DeepTables library, all the model architectures available were tested, including DCN, DeepFM, FGCNN, and PNN. Finally, we implemented the proposed DFG module, which best performed among all other models tested.

\subsubsection*{MRI Temporal Analysis}
Since the dataset contains longitudinal MRI scans, a crucial question is determining the specific time point at which the model excels in accurately diagnosing the dementia subtype. Thus, the experiment was repeated several times, including data at month 0 (baseline), month 3, and month 12 only to assess the MRI scan's importance and model adaptability for early differential diagnosis. 

\subsubsection*{Multi-omics Data Integration}
Finally, since multi-omics data are available, it is essential to investigate if their integration would improve the diagnosis accuracy by comparing results using multi-omics versus MRI only to account for real-life contexts where genetic testing is expensive and not widely available, as can be seen from Figure \ref{fig:feature_improtance}, which investigates the importance of the features. \\

\setlength\tabcolsep{7pt}
\begin{longtable}{@{}cccccc@{}}
\\
\toprule
\thead{Data} &
\thead{Metric} &
\thead{AD Vs\\ CTL} &
\thead{AD Vs\\ MCI} & 
\thead{MCI Vs\\ CTL} &
\thead{AD Vs\\ VaD} 
 \\
    \midrule
\multirow{5}{*}{\textbf{MRI}} & Accuracy & 84.27 & 72.01 & 77.73 & 72.83 \\ 
                              & F1-Score & 80.50 & 70.96 & 77.89 & 69.91 \\
                              & Recall & 79.67& 70.08 & 77.77 &  69.56\\
                              & Precision & 81.34 & 71.87 & 78.02& 70.26\\
                              & AUC &  94.17&  81.05 &  82.54 & 82.36 \\ 
    \bottomrule
\caption{Direct Image Based 3D CNN Results - Table summarizing the accuracy, f1-score, recall, precision, and AUC result of using 3D CNN. The results showcase this limitation compared to the proposed MINDSETS approach.}
\label{CNN_Res}
\end{longtable}

\begin{longtable}{@{}cc|cc|cc|cc|cc|cc@{}}
\toprule
\thead{\textbf{Data}} &
\thead{\textbf{Metric}} &
\multicolumn{2}{c}{\thead{\textbf{AD Vs} \\ \textbf{CTL}}} &
\multicolumn{2}{c}{\thead{\textbf{AD Vs} \\ \textbf{MCI}}} & 
\multicolumn{2}{c}{\thead{\textbf{MCI Vs} \\ \textbf{CTL}}} &
\multicolumn{2}{c}{\thead{\textbf{AD Vs} \\ \textbf{VaD}}} &
\multicolumn{2}{c}{\thead{\textbf{All 4} \\ \textbf{Classes}}} \\
\cmidrule(lr){3-4}\cmidrule(lr){5-6}\cmidrule(lr){7-8}\cmidrule(lr){9-10}\cmidrule(lr){11-12}
& & All & \thead{MRI at \\ time 0} & All & \thead{MRI at \\ time 0} & All & \thead{MRI at \\ time 0} & All & \thead{MRI at \\ time 0} & All & \thead{MRI at \\ time 0} \\
\midrule
\multirow{5}{*}{\rotatebox[origin=c]{90}{\textbf{MRI}}} 
& Accuracy & 97.89 & 98.64 & 82.86 & 87.60 & 83.01 & 83.32 & 82.52 & 87.60 & 52.39 & 62.43 \\ 
& F1-Score & 97.86 & 98.59 & 82.44 & 87.41 & 81.78 & 81.84 & 72.36 & 79.03 & 48.32 & 54.58 \\
& Recall & 98.26 & 98.89 & 82.14 & 88.83 & 80.93 & 83.28 & 76.29 & 82.64 & 58.40 & 61.38 \\
& Precision & 97.65 & 98.33 & 83.20 & 87.25 & 81.95 & 81.04 & 70.35 & 76.61 & 58.55 & 67.64 \\
& AUC & 98.16 & 98.89 & 85.30 & 88.83 & 83.93 & 83.24 & 76.29 & 82.64 & 61.78 & 70.04 \\ 
\midrule
\multirow{5}{*}{\rotatebox[origin=c]{90}{\textbf{Multi-omics}}} 
& Accuracy & 99.35 & 99.92 & 89.52 & \textbf{90.69} & 83.89 & 86.45 & 88.60 & \textbf{89.25} & 70.12 & \textbf{72.23} \\
& F1-Score & 99.31 & 99.92 & 89.27 & \textbf{90.68} & 81.88 & 84.23 & 63.14 & \textbf{81.13} & 70.19 & \textbf{68.29} \\
& Recall & 99.33 & 99.91 & 90.24 & 90.69 & 81.24 & 82.63 & 59.02 & 83.61 & 75.25 & 66.77 \\
& Precision & 99.29 & 99.92 & 88.89 & 90.85 & 82.75 & 87.24 & 70.92 & 79.21 & 67.96 & 75.62 \\
& AUC & 99.33 & 99.91 & 90.69 & 90.24 & 81.24 & 82.63 & 77.07 & 83.61 & 71.26 & 77.32 \\
\bottomrule
\caption{Summary of result using all longitudinal MRI scans, and scan 0 alone versus using it in with other multi-omics data with the DFG module. Results are shown for the binary groups tested and the multiclass experiment.}
\label{MINDSET_DFG}
\end{longtable}

\setlength\tabcolsep{3pt}
\begin{longtable}{@{}ccccccc@{}}
\toprule
\thead{Data} &
\thead{Metric} &
\thead{AD Vs\\ CTL} &
\thead{AD Vs\\ MCI} & 
\thead{MCI Vs\\ CTL} &
\thead{AD Vs\\ VaD} &
\thead{All 4 Classes}
 \\
    \midrule
\multirow{5}{*}{\textbf{MRI}} & Accuracy &  60.66& 67.58 & 69.94 & 61.88 & 44.98 \\ 
& F1-Score & 60.48 & 67.31 &  67.27& 58.90 &  40.64 \\
& Recall & 60.72 & 67.82 &  67.13 & 65.64 &  41.15 \\
& Precision & 60.95 & 68.60 & 67.46 & 76.70 & 49.04  \\
& AUC & 71.45 & 68.60 &  71.45 &  78.29 & 46.25 \\ \hline
\multirow{5}{*}{\textbf{Multi-omics}} & Accuracy  & 81.27 &  80.79&  76.45& 69.72 &  62.91 \\
& F1-Score & 80.92 & 80.05 & 73.28 & 64.35 & 55.34  \\
& Recall & 80.98 &  80.22& 72.66 & 76.94 &  56.69 \\
& Precision & 80.86  & 79.91 &  74.27& 65.85 &  56.29 \\
& AUC & 80.98 & 80.22 & 73.66 &  78.94& 58.54 \\
    \bottomrule
\caption{MINDSETS Approach Without Deep Feature Generation (DFG) Module - Table summarizing the evaluation metric using MRI scans alone versus with multi-omics data using the MINDSETS approach but omitting the DFG module.}
\label{MINDSET_NO_DFG}
\end{longtable}

\section*{Results and Discussion}
\subsection*{Image-Based Baseline} 
The state-of-the-art 3D CNN with DenseNet121 backbone achieved an accuracy of 72.83\% for the classification of AD vs. VaD and an accuracy of 84.27\% for differentiating between AD and CTL, as shown in Table \ref{CNN_Res}. These results showcase the weakness of CNN architecture in discerning the subtle differences in neuroimaging alone and prove the effectiveness of the proposed model.

\subsection*{MINDSETS Approach Results}
Our multi-omics approach integrates radiomics MRI features that were extracted from the segmentation mask of 32 brain structures. The evaluation showcased that 100\% of the 30 randomly reviewed scans were consistent with the radiologist's assessment. Quantitative results with the proposed DFG module are shown in Table \ref{MINDSET_DFG} and without the DFG in Table \ref{MINDSET_NO_DFG} to compare its effect.

\subsubsection*{Qualitative Results}
A significant trend to notice is that based on the result, we can conclude that AD vs. CTL is the easiest group to differentiate, followed equally by AD vs. MCI and MCI vs. CTL, and finally, AD vs. VaD. This trend follows the same pattern as previously published work. In addition, this trend does adhere to the expected medical evaluation wherein it is easier to diagnose between a healthy patient and a patient with full onset AD with higher confidence, followed by the in-between cases where a healthy patient can be showing some cognitive decline that is harder to differentiate from typical aging symptoms or quantifying the measure of the cognitive decline to distinguish between MCI and AD. Since it is established that AD and VaD share a lot of similarities in terms of symptoms, it is expected to be the most challenging class to differentiate. The qualitative aspects of utilizing baseline versus longitudinal MRI data reveal critical insights into the progression and differentiation of dementia subtypes. The baseline MRI provides a snapshot of the brain at a single point in time, which, while helpful, may not capture the dynamic nature of neurodegenerative diseases. In contrast, longitudinal MRI data, encompassing scans from months 0, 3, and 12, offer a temporal dimension that is invaluable for observing the trajectory of the disease. From a qualitative standpoint, the longitudinal approach allows for the detection of subtle changes and trends in brain morphology that could be indicative of early or progressive stages of dementia. This is particularly important in conditions like MCI, which may show minimal changes at month 0 but more pronounced changes by month 12. The inclusion of data from multiple time points improves the model's ability to classify conditions like AD and VaD, where the rate and pattern of progression can vary significantly between individuals. Finally, when comparing the results with and without the proposed DFG module, we can see a clear trend where the performance metrics across all categories show enhancement when the DFG module is used. The trend is consistent across other classifications, with the most substantial gains evident in the binary classifications.

\subsubsection*{Quantitative Results}
The radiomics features, genetic, and clinical data achieved an accuracy of 99.35\% for AD vs. CTL and 88.60\% for AD vs. VaD using multi-omics data and all longitudinal scans. This method significantly surpasses traditional techniques, especially when differentiating AD from MCI (89.52\%) and multiclass scenarios (70.12\%). Baseline MRI scans also show robust accuracy (89.25\% for AD vs. VaD), emphasizing early diagnostic potential. This result has critical implications as it stresses the importance of early intervention to improve diagnostic accuracy. Multi-omics data are proven essential, enhancing accuracy by an average of 6.5\% 

These enhancements suggest that the model has captured the intricate patterns within the multi-omics data, enabling it to navigate the subtle distinctions between dementia subtypes that share symptomatic and pathological overlap. Moreover, in the case of AD vs. VaD, as established, AD patients perform worse on memory tasks than VaD; therefore, the MMSE scores help further consolidate this difference when classifying between the two types. However, it is equally important to consider the performance using MRI scans alone since other data may be expensive to acquire or unavailable. The most significant boost witnessed was for the AD vs. MCI groups, highlighting the challenge of diagnosing using only morphological brain changes detectable by MRI. Finally, it is clear that the proposed DFG module boosts the performance as in the binary classification of AD vs CTL using MRI data, accuracy jumps from 60.66\% without DFG to 97.89\% with DFG. Similarly, multi-omics data performance increases significantly with the addition of DFG, as seen in the same classification, where accuracy rises from 81.27\% to 99.35\%.

\subsection*{Treatment Effect Assessment and Evaluation}
\begin{figure}
\centering
\includegraphics[width=0.6\textwidth]{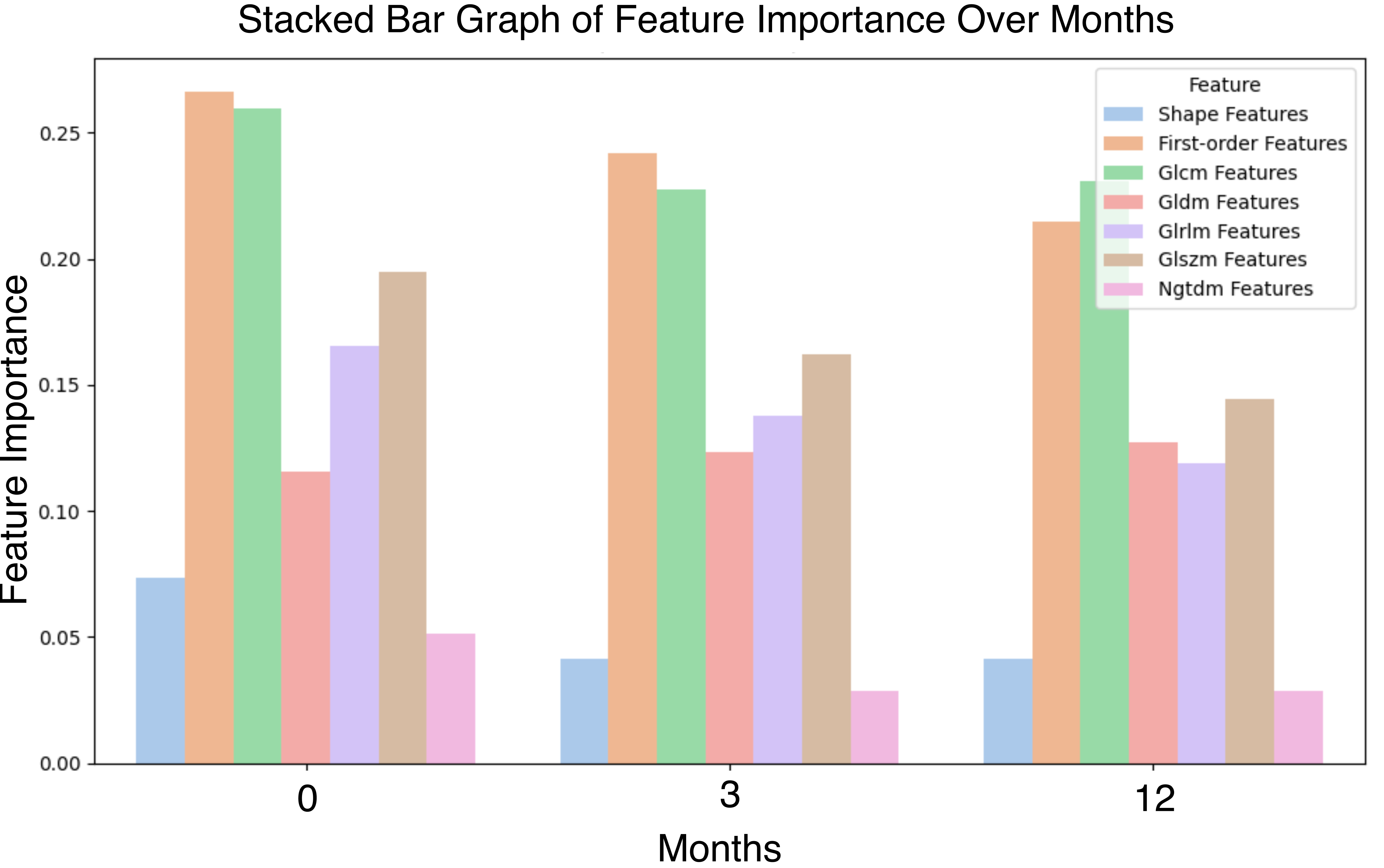}
\caption{Feature importance of radiomics MRI features for treated MCI patients, showcasing a notable decrease in the relative importance as time progresses.} \label{treat_FI}
\end{figure}

To investigate the effect of treatment on dementia progression, a subset of 20 MCI patients was compared to a control group of 20 MCI patients who were not on medication. Firstly, the comparison is broken down in terms of MRI scan timepoint. The softmax probability of the patient belonging to the MCI class was then observed. It is important to note that both classes (CTL and MCI) were balanced. Therefore, a decrease in the probability of belonging to the MCI class is correlated with an improvement or a slowing of progression for the patient. The change was recorded after 3 months as well as after 12 months. Over 3 months, 15 out of 20 patients on medication had a decrease of probability belonging to the MCI class with an average of 5.35\% decrease. In the control group who were not on medication, only 5 patients witnessed a decrease with an average of 2.66\%. Over the 12 months, 17 out of the 20 patients on medication had a decrease of probability belonging to MCI class with an average of 6.38\% decrease; for the control group, only 6 had a decrease of an average of 2.53\%. Moreover, when analyzing the feature importance for the MRI radiomics features over time for the medicated patients, there is a downward trend where the feature importance decreases as time goes on, as shown in Figure \ref{treat_FI}. Overall, this data suggests that medication may play a role in slowing the progression of MCI, but further research is necessary to confirm these findings and to understand the broader implications for patient care.

\subsection*{Clinical Significance}
Based on the results achieved, the proposed MINDSETS approach offers several significant advancements in the early diagnosis and treatment planning for AD and VaD. Integrating multi-omics data with neuroimaging provides a comprehensive diagnostic framework that improves accuracy and offers deeper insights into the underlying pathophysiological differences between AD and VaD. 

\subsubsection*{Impact on Early Diagnosis}
Early and accurate diagnosis of dementia subtypes is critical for effective patient management and treatment planning. Traditional diagnostic methods often struggle with the overlapping symptoms of AD and VaD, leading to delayed or incorrect diagnoses. The MINDSETS approach, with its diagnostic accuracy of 88.60\% for AD vs. VaD, represents a significant improvement over existing methods. This high accuracy ensures that patients receive the correct diagnosis sooner, which is crucial for initiating appropriate treatment strategies.

\subsubsection*{Clinical Implications}
\begin{itemize}
    \item \textbf{Enhanced Diagnostic Precision:} By distinguishing between AD and VaD with high accuracy, our model can reduce misdiagnosis rates, ensuring that patients receive condition-specific treatments early in the disease course.
    \item \textbf{Targeted Treatment Plans:} Accurate differentiation allows clinicians to tailor treatment plans that are specific to the type of dementia. For example, while AD treatments may focus on slowing neurodegeneration and managing symptoms, VaD treatments would emphasize managing cardiovascular risk factors to prevent further vascular damage.
    \item \textbf{Improved Patient Outcomes:} Early and precise diagnosis can lead to better patient outcomes by slowing disease progression, reducing symptom severity, and improving quality of life.
\end{itemize}

\subsubsection*{Implications for Treatment Planning}
The ability of the MINDSETS model to monitor longitudinal changes and assess treatment efficacy is another key contribution. By tracking disease progression over time through MRI and multi-omics data, the model can help clinicians evaluate the effectiveness of treatments and adjust them as needed.

\noindent \textbf{Clinical Practice Guidelines:}
\begin{itemize}
\item \textbf{Personalized Medicine:} The explainable nature of the model, facilitated by feature importance scores, supports personalized medicine by identifying the most relevant biomarkers for each patient. This can guide clinicians in selecting the most effective treatment options.
\item \textbf{Monitoring and Adjusting Treatments:} The model's longitudinal aspect enables continuous patient monitoring, providing critical information on how well a treatment is working and whether adjustments are necessary. This aligns with clinical practice guidelines that recommend regular monitoring and adjustment of treatment plans based on patient response.
\end{itemize}

\section*{Conclusion}
In conclusion, the research conducted using the model has provided significant insights into the capabilities of machine learning in the differential diagnosis of AD versus VaD. By comparing baseline and longitudinal MRI data, enhanced with clinical, radiomics, genetic, and cognitive assessments, we have observed that integrating multi-omics data contributes to a more accurate and nuanced classification of dementia. Using the MRI scans alone without additional modalities provides good results, an excellent option for reducing testing costs. The diagnostic accuracy in distinguishing between AD and VaD peaks at 89.25\% using the multi-omics data. The investigation also revealed the effectiveness of the proposed MINDSETS approach with the DFG module in discerning the subtle differences due to its advanced feature generation capabilities. Another significant finding indicates the need for longitudinal MRI scans when differentiating and tracking progressive neurodegenerative disease, especially in the MCI group. The reason behind this is that it provides temporal insight, which is particularly valuable in complex cases. The DFG model's performance in utilizing all available longitudinal scans suggests that the trajectory of neurodegenerative changes is an essential factor in the disease classification process. Notably, in the multifaceted task of distinguishing between dementia subtypes, the longitudinal data approach demonstrated a marked improvement in model accuracy and predictive capability.

\noindent\textbf{\\\\Future Directions}
A critical future direction of the model is to ensure the model's generalizability. Future research should aim to validate findings across different populations and healthcare settings. This will help to mitigate biases and ensure that diagnostic tools are practical across diverse demographic groups. This direction is essential in this case as it is established that demographics play a vital key in neurodegenerative diseases, which can help in diagnosis.

\bibliography{sample}

\section*{Competing interests}
The authors declare no competing interests.

\section*{Data availability}
The data supporting the findings of this study is ANMerge, which is openly available and can be accessed via Synapse at \url{https://doi.org/10.7303/syn22252881.}
The code developed for the analysis and implementation will be publicly available upon accepting the manuscript.

\end{document}